\documentclass[12pt,amstex,aps,amsfonts,prsty]{article} 
\usepackage[dvips]{graphicx}
\usepackage{amssymb}
\usepackage{amsmath}
\usepackage{a4wide}
\usepackage{citesort}

\begin{document}

\begin{titlepage}

\begin{center}

{\Huge \bf Intrinsic friction of adsorbed monolayers}

\vspace{0.3in}

{\Large \bf O.B{\'e}nichou$^{1}$, A.M.Cazabat$^1$,  J.De Coninck$^{2}$,
 M.Moreau$^3$ and  G.Oshanin$^3$}

\vspace{0.1in}

{\large \sl $^1$ Laboratoire de Physique de la Mati{\`e}re Condens{\'e}e, \\
Coll{\`e}ge de France, 11 Place M.Berthelot, 75252 Paris Cedex 05, France
}

{\large \sl $^2$ Centre de Recherche en Mod{\'e}lisation Mol{\'e}culaire, \\
Universit{\'e} de Mons-Hainaut, 20 Place du Parc, 7000 Mons, Belgium
}

{\large \sl $^3$ Laboratoire de Physique Th{\'e}orique des Liquides, \\
Universit{\'e} Paris 6, 4 Place Jussieu, 75252 Paris, France
}

\begin{abstract}

In the present paper  we overview our recent results on
 intrinsic frictional properties
 of adsorbed monolayers,
composed of mobile hard-core particles  
undergoing continuous exchanges with a vapor phase. 
Within the framework of a dynamical master equation approach, describing the time
 evolution of the system, we determine in the most general form
 the terminal velocity of 
some biased impure molecule -
the tracer particle (TP), 
constrained to move inside the adsorbed monolayer probing its
frictional properties, 
define the frictional forces
as well as the
particles density distribution in the monolayer.
Results for one-dimensional solid substrates, appropriate to adsorbtion on polymer
chains, are compared against the Monte Carlo
simulation data, which confirms our analytical predictions.  

\end{abstract}

\end{center}

\end{titlepage}


\section{Introduction.}

In the present paper we overview our recent results on
intrinsic friction of monolayers
emerging on solids exposed to 
a vapor phase \cite{benichou,physa,prl,benichouPRB}.
Such layers are involved in various 
technological and material processing operations, including, for instance, 
coating, gluing or lubrication. Knowledge of their intrinsic frictional properties is
important for conceptual understanding of different transport processes taking place 
within molecular films, film's stability, 
as well as spreading of ultrathin 
liquid films on solid surfaces \cite{spreading,spreading1}, 
spontaneous or forced dewetting of monolayers 
\cite{aussere,dewetting,dewetting2,persson}
or island formation on solid surfaces \cite{islands}.

Since the early works of Langmuir, 
much effort has been invested in the analysis of the
equilibrium properties of
the adsorbed films
\cite{surf,shaw,fowler,1,2}. Here,  
significant analytical results have been obtained 
predicting  different phase transitions and ordering 
phenomena, which well agree with the available experimental data. 
As well, 
some approximate results have been obtained for
both dynamics of an isolated adatom
on a corrugated surface and collective
diffusion, describing spreading of the macroscopic 
density fluctuations in
 interacting adsorbates being in contact with 
the vapor 
\cite{gomer,kreuzer,gortel,wahn}. 

Another important aspect of dynamical behavior concerns tracer
 diffusion in adsorbates,
which is observed experimentally in  
STM or field ion measurements
 and provides
a useful information about adsorbate's viscosity or intrinsic
friction. This problem is not only a challenging question
in its own right due to emerging non-trivial, essentially cooperative behavior, 
but is also
crucial for understanding of various dynamical processes taking
place on solid surfaces. However, most of available theoretical 
studies of tracer diffusion in adsorbed layers 
(see, e.g.
Refs.\cite{kehr,nakazato,tahir,beijeren,hilhorst,bur1,bur2,oshanin,deconinck}) 
do exclude the
 possibility of
particles exchanges with the vapor.

Here we focus on this important issue and provide a theoretical description
of the properties of tracer diffusion in
adsorbed monolayers in contact with a vapor phase - a reservoir of particles. 
More specifically, the system we consider 
consists of (a)
a solid substrate, which is modeled
in a usual fashion as a regular
 lattice of adsorbtion sites; (b)  
a monolayer of adsorbed, mobile
hard-core particles in contact with a vapor 
and (c) a single hard-core tracer
particle (TP). We suppose that the monolayer particles move randomly
along the lattice
by performing symmetric
hopping motion between
the neighboring lattice sites, which process
 is constrained by mutual hard-core interactions, and 
 may desorb from and adsorb onto the lattice from the vapor
with  some prescribed rates dependent on the vapor 
pressure, temperature
 and the interactions with the solid substrate.
In contrast, the
tracer particle is constrained to move 
along the 
lattice only, 
(i.e. it can not desorb 
to the vapor), and  is
subject to
a constant external force of an arbitrary magnitude $E$. Hence, 
the TP performs
a biased random walk,  
constrained by the  hard-core
interactions with the
monolayer particles, 
and always remains within 
the monolayer, probing 
its frictional properties.

The questions  we address here are the following: First, focussing on one- and
two-dimensional systems,  
we aim to determine 
the  force-velocity
relation. 
That is, as $t \to \infty$, the TP ultimately attains a constant velocity $V_{tr}(E)$, which depends on the magnitude of the
applied external force; the functional form of this dependence in the most general case
constitutes the primary goal of our analysis.   
Next, we study the form of the force-velocity  relation in the limit of a vanishingly small external bias. 
This allows us, in particular, to show that the frictional force exerted on the TP by the monolayer
particles is viscous, and
to evaluate the corresponding
friction
coefficient. Lastly, we analyse how the biased TP perturbs the particles density distribution in the
monolayer. As a matter of fact, we proceed to show  
that there are stationary density profiles around the TP as $t \to \infty$, which mirror a remarkable cooperative
behavior.   

We finally remark that our model can be viewed from a somewhat different perspective. 
Namely, on the one hand, the model under study
is  a certain generalization of the  ``tracer diffusion in
a  
hard-core lattice gas'' problem (see, e.g. Ref.\cite{kehr} for an extensive review) to the case where the random walk
performed by the TP is {\it biased} and the number of particles in
the monolayer is {\it not explicitly conserved}, due to exchanges with
the reservoir. We recall that even this, by now classic model
constitutes a many-body problem for which no exact general expression of
the tracer diffusion coefficient $D_{tr}$ is known. 
On the other hand, our model represents a novel example 
 of the so called ``dynamical percolation'' models \cite{drugerA,drugerB,zwanzig,sahimi,chatterjee,nouspercol},
invoked to describe transport processes in
 many situations with dynamical disorder. In this context,
the particles of the monolayer can be thought of as representing
some fluctuating
environment, which hinders the motion of an impure molecule - the TP, which might be, for example, a charge carrier. 
An important aspect  of our model, 
which makes it different from the previously proposed models of dynamic
percolative environments, is that we include the hard-core 
interaction between "environment" particles and the 
tracer particle, such that the latter may influence itself the dynamics of the 
environment.  
Lastly, we note that the model under study can be thought of as 
some simplified picture of the 
stagnant layers emerging in liquids being in contact with a solid body. It is well
known (see, e.g. Ref.\cite{lyklema}) that liquids in close vicinity
 of a solid interfaces - at distances within a few molecular diameters, do possess completely 
different physical properties compared to these of the bulk phase.
In this "stagnant" region, in which an intrinsically disordered liquid phase is spanned by and contends
with the ordering potential of the solid, 
liquid's viscosity is drastically enhanced 
and transport processes, (regarding, 
say, biased diffusion of charged carriers in solutions), are
essentially hindered. Thus our model can be viewed 
as a two-level approximate model of this challenging physical system,
in which
the reservoir mimics the bulk fluid phase with very rapid transport,
while the adsorbed monolayer represents the stagnant layer emerging on the solid-liquid interface.

The paper is structured as follows: In Section \ref{generalpresentation}
  we formulate
our model in case of a  two-dimensional solid substrate and introduce
basic notations.  We write down then the  dynamical
equations which govern the time evolution of the monolayer
particles and of the tracer, and outline the decoupling approximation used
to close this system of equations. These equations are presented for the general, $d$-dimensional case. 
Sections  \ref{1D}  and \ref{2D}  are respectively  devoted to the presentation of
the results of this general approach for one-dimensional substrates, which situation is appropriate to adsorbtion on polymer
chains \cite{10,11}, and 
also for two-dimensional solid substrates.
Finally, we conclude in Section \ref{Conclusion}
  with a brief summary
and discussion of our results.

\section{The model and basic equations}\label{generalpresentation}

\subsection{The model.}

\begin{figure}[h]
\begin{center}
\includegraphics*[scale=0.4]{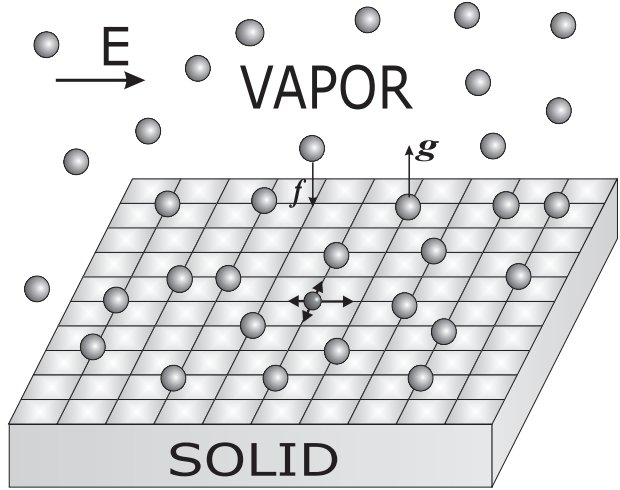}
\end{center}
\caption{\label{reseau}Adsorbed monolayer in contact with a vapor. 
Grey spheres denote the
monolayer (vapor) particles; the smaller black sphere 
stands for the driven tracer
particle.}
\end{figure}

Consider a
two-dimensional 
solid surface with some concentration of adsorbtion sites,
which is brought 
in
contact with a reservoir containing identic,
electrically neutral particles  - a vapor phase (Fig.\ref{reseau}), 
maintained at a constant pressure. For simplicity of exposition, we assume here
that adsorbtion sites form a regular square lattice
of spacing $\sigma$.  
We suppose next that the reservoir particles 
may adsorb 
onto any vacant adsorbtion site at a fixed rate $f/\tau^*$, which rate 
depends on the vapor
pressure and the energy gain due to the adsorbtion event. Further on,
the adsorbed particles may move randomly along the lattice by  
hopping at a rate $1/ 4 \tau^*$ to any of $4$
neighboring adsorbtion sites,
which process is 
constrained by hard-core exclusion preventing multiple occupancy of any of the sites.
Lastly, the adsorbed particles may 
 desorb from the lattice  back to the reservoir
at rate $g/\tau^*$, which is dependent on the barrier against desorption. 
Both $f$ and $g$ are site and environment independent.

Note that such a model represents, of course, a certain idealization. 
In "real world" systems, 
the adsorbed particles experience two 
types of interactions: 
the solid-particle (SP) interactions, i.e. interactions  
with the atoms of the host solid,
and the particle-particle (PP) interactions, i.e. mutual interactions with each other. 
Now, both the SP and the PP  interactions  
are characterized by a harsh repulsion at short scales,  
and are attractive at longer distances. The SP repulsion
 keeps the adsorbed particles 
some distance apart of the solid, while the SP attraction
favors adsorbtion and   
hinders particles desorption as well as migration along the solid surface. 
In this regard, our model corresponds to the regime of the so-called
intermediate localized adsorption \cite{1,2}: the particles forming an adsorbed monolayer are neither completely fixed
in the potential wells created by the SP 
interactions, nor completely mobile.
This means, that the potential wells are rather deep with 
respect to particles desorption (desorption barrier $U_d \gg k T$), such that the adsorbate forms a 
submonolayer, 
but nonetheless, have a much lower energy barrier $V_l$ against the lateral movement across the
surface,  $U_d \gg V_l > k T$. 
In this regime, each adsorbed particle spends a considerable part of its time at the bottom
of a potential well and jumps sometimes, solely due to the thermal activation, from one potential minimum to another
in its neighborhood; after the jump is performed, the particle dissipates all its energy to the host solid. 
Thus, on a macroscopic time scale 
the particles do not possess any velocity. 
The time $\tau^*$ separating two
successive jump events, is just 
the typical  time a given particle 
spends in a given well 
vibrating around its minimum;  $\tau^*$ 
is related to the temperature, the barrier for 
lateral motion and the frequency of the
solid atoms' vibrations by the  Arrhenius formula.

We emphasize that such a type of random 
motion is essentially different of the standard hydrodynamic picture of particles
random motion in the two-dimensional "bulk" liquid phase, e.g. in 
free-standing liquid films, 
in which case there is a
velocity distribution and spatially $random$ motion results from the PP scattering. 
In this case,
the dynamics may  be only approximately considered as an activated hopping of particles, confined to some effective
cells by the potential field of their neighbors, along a lattice-like structure of such cells (see, e.g.
Refs.\cite{3,4}). In contrast to the dynamical model to be studied here, standard two-dimensional hydrodynamics
pressumes that the particles do not interact with the underlying solid. In 
realistic systems, of course, both the
particle-particle scattering and scattering by the potential wells due to the interactions with the host solid, (as
well as the corresponding dissipation channels), are  important \cite{kreuzer,6}. 
In particular, it has been shown that addition of a dissipation to the host solid removes
the infrared divergencies in the dynamic density correlation 
functions and thus makes the transport coefficients finite 
\cite{7,8}. 
On the other hand, homogeneous adsorbed monolayers may only exist
 in systems in which the
attractive part of the PP
interaction potential is essentially  weaker than that
 describing interactions with the solid; otherwise, such
monolayers become unstable and dewet spontatneously from the solid surface.
As a matter of fact, for stable homogeneous monolayers,  
the PP interactions are at least ten times weaker that the
interactions with the solid atoms \cite{2}. 

Consequently, the standard hydrodynamic picture of 
particles dynamics is
inappropriate 
under the defined above physical conditions. Contrary to that, 
any adsorbed particle moves due to random hopping events, 
activated by chaotic vibrations
 of the solid
atoms, along the local minima of an array of potential wells, 
created due to the interactions with the solid
\cite{1,2}. As we have already remarked, in the physical conditions 
under which such a dynamics takes place, 
the PP interactions are much weaker than the SP interactions
and hence do not perturb significantly
the regular array of potential wells due to the SP interactions. In our model, 
we discard completely the attractive part of
the PP interaction potential and take into account only the repulsive one, 
which is approximated by an abrupt, hard-core-type
potential.    
 
To describe the occupancy of lattice sites, we introduce 
a time-dependent 
variable $\eta({\bf R})$, which may assume two values:
\begin{equation}
\eta({\bf R}) = \left\{\begin{array}{ll}
1,     \mbox{ if the site ${\bf R}$ is occupied by an adsorbed particle,} \nonumber\\
0,     \mbox{ if the site ${\bf R}$  is empty.}
\end{array}
\right.
\end{equation} 
Note that the local variable $\eta({\bf R})$ can change its value due to adsorption,
desorption and random hopping events. Note also that random hopping events do conserve the total number of adsorbed
particles and hence, the average adsorbate density $\rho_s(t)$. On the other hands, 
adsorption/desorption processes do change $\eta({\bf R})$ locally such that   
the total number of 
particles in the adsorbed monolayer 
is not explicitly conserved. 
However, the mean density
of the adsorbate, $\rho_s(t) = <\eta({\bf R})>$, approaches as
$t\to\infty$ a constant value
\begin{equation}
\rho_s=\frac{f}{f+g}
\label{rhostat}
\end{equation} 
This  relation is well known and represents the customary Langmuir adsorption
isotherm \cite{surf}. We finally remark that in the analysis of the stationary-state behavior,
 we can always turn to the conserved particles number limit by 
setting $f$ and $g$ equal to zero
 and keeping their ratio
fixed, i.e. supposing that $f/g = \rho_s/(1 - \rho_s)$. 
This limit will correspond to the model of biased tracer diffusion in
a  hard-core lattice gas with fixed particles density $\rho_s$, 
and will allow us to check our analytical
predictions against some already known results
\cite{kehr,nakazato,tahir,beijeren,hilhorst,deconinck}. 
We will furnish such a comparison at the end of
the section \ref{2D}.

Further on, at $t = 0$ we introduce at the lattice origin 
an extra hard-core 
particle, whose motion
we would like to follow;  
position of this particle at time $t$ is denoted as ${\bf R}_{tr}$. Note that the tracer particle (TP) is
designed to
measure the resistance offered by the monolayer
particles to the external perturbance, or, in other words, 
to measure the intrinsic frictional properties of the adsorbate. 

Now, we stipulate that the TP is different from the adsorbed particles in two aspects:
first, it can not desorb from the lattice and second,  
it is subject to
some external driving force, which favors its jumps into a
preferential direction. 
Physically, such a  situation may be realized, 
for instance,
if this only particle is charged and the system is subject to a
uniform electric field {\bf E}.  We suppose here, 
for simplicity of exposition, that the external force ${\bf E}$
 is oriented according to the unit vector 
${\bf e_1}$.

The dynamics of the biased TP is defined in the usual fashion:
We suppose
that the TP, which occupies the site ${\bf R}_{tr}$ at time $t$,
 waits an exponentially distributed time with mean
$\tau$, 
and then attempts to hop onto one of $4$ neighboring sites, ${\bf
R}_{tr} + {\bf e}_{\boldsymbol \nu}$, where ${\bf e}_{\boldsymbol \nu}$
are $4$  unit 
vectors of the hypercubic lattice. 
In what follows we adopt the notation $\nu = \{\pm1,\pm2\}$, where 
${\bf \pm e_1}$  will denote 
the  direction of the external force ${\bf E}$.  Next, the jump direction is chosen 
according to the probablity $p_\nu$, which obeys:
\begin{equation}
p_\nu=\frac{\exp\Big[\frac{\beta}{2}({\bf E \cdot e}_{\boldsymbol
\nu})\Big]}{\sum_{\mu}\exp\Big[\frac{\beta}{2}({\bf E \cdot e}_{\boldsymbol \mu})\Big]},
\label{defp}
\end{equation}    
where $\beta$ is the reciprocal temperature, $({\bf E \cdot e})$ stands for the scalar product, 
the charge of the TP is set equal to unity and  
the sum with the subscript $\mu$ denotes summation over all possible
orientations
of the vector ${\boldsymbol e_\mu}$; that is, $\mu = \{\pm1,\pm2\}$.

After the jump direction is chosen, the TP attempts to hop 
onto the target site. The hop is instantaneously fulfilled 
if 
the target site
is vacant at this moment of time; otherwise, i.e., if the target site
is occupied by any adsorbed particle, the jump is rejected and the
TP remains at its position.

\subsection{Evolution equations.}

Now, we derive the evolution equations in a general, $d$-dimensional case, which will allow us to compare the behavior
emerging in one- and two-dimensional systems.   We begin by introducing some auxiliary definitions.
Let $\eta\equiv
\{\eta({\bf R})\}$ denote the entire set of the occupation variables, which defines the instantaneous
configuration of the adsorbed particles  at the 
lattice at time moment $t$. Next, let 
$P({\bf R_{tr}},\eta;t)$ stand for the joint probability of finding  
at time $t$ the TP at the site ${\bf R_{tr}}$
and all adsorbed particles in the configuration $\eta$.
Then, denoting as $\eta^{{\bf r},\nu}$ a
configuration obtained from $\eta$ by the Kawasaki-type exchange of the occupation
variables of two neighboring 
sites ${\bf r}$ and ${\bf r+e}_{\boldsymbol \nu}$,
and as 
${\hat \eta}^{\bf r}$ - a configuration obtained from 
the original  $\eta$ by
the replacement $\eta({\bf r}) \to 1-\eta({\bf r})$, which corresponds to the
Glauber-type flip of the occupation variable due to 
the adsorption/desorption 
events, we have that 
the time evolution of the configuration
probability $P({\bf R_{tr}},\eta;t)$ obeys the following master equation:
\begin{eqnarray}
&&\partial_tP({\bf R_{tr}},\eta;t)=
\frac{1}{2d\tau^*}\sum_{\mu=1}^d\;\sum_{{\bf r}\neq{\bf R_{tr}}-{\bf e}_{\boldsymbol \mu},{\bf R_{tr}}}  \; 
\Big\{ P({\bf R_{tr}},\eta^{{\bf r},\mu};t)-P({\bf R_{tr}},\eta;t)\Big\}\nonumber\\
&+&\frac{1}{\tau}\sum_{\mu}p_\mu\Big\{\left(1-\eta({\bf R_{tr}})\right)P({\bf R_{tr}}-{\bf e}_{\boldsymbol \mu},\eta;t)
-\left(1-\eta({\bf R_{tr}}+{\bf e}_{\boldsymbol \mu})\right)P({\bf R_{tr}},\eta;t)\Big\}\nonumber\\
&+&\frac{g}{\tau^*}\sum_{{\bf r}\neq {\bf R_{tr}}} \;\Big\{\left(1-\eta({\bf r})\right)P({\bf R_{tr}},
\hat{\eta}^{{\bf r}};t)-\eta({\bf r})P({\bf R_{tr}},\eta;t)\Big\}\nonumber\\
&+&\frac{f}{\tau^*}\sum_{{\bf r}\neq{\bf R_{tr}}} \;\Big\{\eta({\bf r})P({\bf R_{tr}},\hat{\eta}^{{\bf r}};t)
-\left(1-\eta({\bf r})\right)P({\bf R_{tr}},\eta;t)\Big\}.
\label{eqmaitresse}
\end{eqnarray} 

The mean 
velocity $V_{tr}(t)$ 
of the TP can be obtained by multiplying both sides of Eq.(\ref{eqmaitresse}) by
$({\bf R_{tr} \cdot e_1})$ and summing over all possible configurations $({\bf R_{tr}},\eta)$.
This results in the following exact equation determining the TP velocity:
\begin{equation}
V_{tr}(t)\equiv\frac{d}{dt} \; \sum_{{\bf R_{tr}},\eta} ({\bf R_{tr} \cdot e_1})P({\bf R_{tr}},\eta;t) =
\frac{\sigma}{\tau}\Big\{p_1  \Big(1-k({\bf e_1};t)\Big)-p_{-1} \Big(1-k({\bf e_{-1}};t)\Big)\Big\},
\label{vitesse}
\end{equation}
where 
\begin{equation}
k({\boldsymbol \lambda};t)\equiv\sum_{{\bf R_{tr}},\eta}\eta({\bf R_{tr}}+{\boldsymbol \lambda})P({\bf R_{tr}},\eta;t)
\label{defk}
\end{equation}
is the probability of having at time t an 
adsorbed particle 
at position ${\boldsymbol \lambda}$, 
defined in the frame of reference moving with the TP. 
In other words,  $k({\boldsymbol \lambda};t)$
 can be thought of as being 
the density
profile in the adsorbed monolayer as seen from the moving TP. 

Equation (\ref{vitesse}) signifies that 
the velocity of the TP is 
dependent on the monolayer particles density in the 
immediate vicinity of the tracer. 
If the monolayer is perfectly stirred, or, in other words, if 
$k({\boldsymbol \lambda};t) = \rho_s$ everywhere, (which implies immediate 
 decoupling of ${\bf R_{tr}}$ and $\eta$),  one would obtain
from Eq.(\ref{vitesse}) a trivial mean-field result
\begin{equation}
V_{tr}^{(0)}=(p_1-p_{-1})(1-\rho_s)\frac{\sigma}{\tau},
\label{vmf}
\end{equation}
which states that the only effect of the medium on the TP dynamics is that its 
jump time $\tau$ is merely renormalized by a
 factor $(1 - \rho_s)^{-1}$, which represents the inverse 
concentration of voids in the monolayer; note that then 
$(1 - \rho_s)/\tau $ defines simply 
the mean frequency of successful jump events. 

However, the situation appears to be more complicated and, as we proceed to show, 
$k({\boldsymbol \lambda};t)$ is 
different from the equilibrium value $\rho_s$ everywhere, except for 
 $|\boldsymbol \lambda|\to\infty$. This means that the TP strongly 
perturbs the particles distribution in the monolayer - it is no longer
uniform and some non-trivial stationary density profiles emerge.

Now, in order to calculate the instantaneous mean
velocity of the TP we have to determine the mean particles density 
at the neighboring to the TP sites
${\bf R_{tr}}+{\bf e_{\pm1}}$, which requires, in turn, computation of 
the density profile $k({\boldsymbol \lambda};t)$ for arbitrary $\boldsymbol \lambda$.
The latter can be found
 from the master equation     
(\ref{eqmaitresse}) by multiplying both sides 
by $\eta({\bf R_{tr}})$ and performing the summation over all
configurations $({\bf R_{tr}},\eta)$. In doing so, we find 
that these equations are not closed with respect to 
$k({\boldsymbol \lambda};t)$, 
but are coupled to the third-order
correlations,
\begin{equation}
T({\boldsymbol \lambda},{\bf e}_{\boldsymbol \nu};t) =\sum_{{\bf R_{tr}},\eta}\eta({\bf R_{tr}}+{\boldsymbol \lambda})\eta({\bf R_{tr}}+{\bf e}_{\boldsymbol \mu})P({\bf R_{tr}},\eta;t)
\end{equation}
In turn, if we proceed further to
the third-order
correlations, we find that these are 
coupled respectively to the fourth-order correlations. 
Consequently, in order
to compute $V_{tr}$, one faces
the problem of solving an infinite hierarchy of coupled equations
for the correlation functions.  
Here we resort to the simplest non-trivial 
closure of the hierarchy    
in terms of $k({\boldsymbol \lambda};t)$, which has been first proposed in
Refs.\cite{bur1} and \cite{bur2}, and  represent $T({\boldsymbol
\lambda},{\bf e}_{\boldsymbol \nu};t)$ 
as
\begin{eqnarray}
&&\sum_{{\bf R_{tr}},\eta}\eta({\bf R_{tr}}+{\boldsymbol \lambda})\eta({\bf R_{tr}}+{\bf e}_{\boldsymbol \mu})P({\bf R_{tr}},\eta;t)\nonumber\\
&\approx&\left(\sum_{{\bf R_{tr}},\eta}\eta({\bf R_{tr}}+{\boldsymbol \lambda})P({\bf R_{tr}},\eta;t)\right)\left(\sum_{{\bf R_{tr}},\eta}
\eta({\bf R_{tr}}+{\bf e}_{\boldsymbol \mu})P({\bf R_{tr}},\eta;t)\right)\nonumber\\
&=&k({\boldsymbol \lambda};t)k({\bf e}_{\boldsymbol \mu};t), 
\label{decouplage}
\end{eqnarray}
Some arguments justifying such an approximation {\it a posteriori} are
presented in Sections 3 and 4 (see also  Ref.\cite{benichou}).

Using  the approximation in Eq.(\ref{decouplage}), we obtain
\begin{equation}
2d\tau^*\partial_tk({\boldsymbol \lambda};t)=\tilde{L}k({\boldsymbol \lambda};t)+2df, 
\label{systemek1}
\end{equation}
which holds for all ${\boldsymbol \lambda}$, except for  ${\boldsymbol \lambda}=\{{\bf 0},\pm{\bf e_1},{\bf e}_{2}\ldots,{\bf e_d}\}$. One the
other hand, for these special sites ${\boldsymbol \lambda} = {\bf e_{\nu}}$ 
with $\nu=\{\pm1,2,\ldots, d\}$ we find
\begin{equation}
2d\tau^*\partial_tk({\bf e}_{\boldsymbol \nu};t)=(\tilde{L}+A_\nu)k({\bf e}_{\boldsymbol \nu};t)+4f,  
\label{systemek2}
\end{equation}
where 
$\tilde{L}$ is the operator  
\begin{equation}
\tilde{L}\equiv\sum_\mu A_\mu\nabla_\mu-4(f+g),
\end{equation}
and the coefficients $A_{\mu}$ are defined by
\begin{equation}
A_\mu(t)\equiv1+\frac{2d\tau^*}{\tau}p_\mu(1-k({\bf e}_{\boldsymbol \mu};t)).
\label{defA}
\end{equation}
Note that Eq.(\ref{systemek2}) represents, from the mathematical point of view, 
the boundary conditions for the general evolution equation  (\ref{systemek1}), imposed on the sites in the
immediate vicinity of the TP. Equations (\ref{systemek1}) and (\ref{systemek2}) together with Eq.(\ref{vitesse}) thus consitute
a closed system of equations which suffice computation of all properties of interest.

\subsection{Stationary solution of the evolution equations}\label{methode}

We turn to the limit $t\to\infty$  and suppose that both the density profiles and stationary velocity of the TP
 have non-trivial stationary values
  \begin{equation}
  k({\boldsymbol \lambda})\equiv\lim_{t\to\infty}k({\boldsymbol \lambda};t), \;\;\;
  V_{tr}\equiv\lim_{t\to\infty}V_{tr}(t),\;\;\;\mbox{and}\;\;\; A_\mu \equiv \lim_{t\to\infty}A_\mu(t)
  \end{equation}
Define next
the local deviations of $k({\boldsymbol \lambda})$ from the unperturbed density as
  \begin{equation}
  h({\boldsymbol \lambda})\equiv k({\boldsymbol \lambda})-\rho_s
  \label{defh}
  \end{equation}
Choosing that $h({\bf 0})=0$, we obtain then the following fundamental
system of equations: 
  \begin{equation}
\tilde{L}h({\boldsymbol \lambda})=0\;\;\;\mbox{for \; all}\;\;\; 
{\boldsymbol \lambda} \neq \{{\bf 0},{\bf e_{\pm 1}},\ldots, {\bf e_{\pm d}}\},
  \label{systemeh1}
  \end{equation}
  \begin{equation}
 (\tilde{L}+A_\nu)h({\bf e}_{\boldsymbol \nu})+\rho_s(A_\nu-A_{-\nu})=0\;\;\;\mbox{for}\;\;\; 
{\boldsymbol \lambda} = \{{\bf 0},{\bf e_{\pm 1}},\ldots, {\bf e_{\pm d}}\},
  \label{systemeh2}
  \end{equation}
which determine the deviation from the unperturbed density $\rho_s$ 
in the stationary state.  Note also that in virtue of an evident symmetry,   
$h({\bf e_{\nu}}) =h({\bf e_{-\nu}})$
and $A_\nu  = A_{-\nu} $ for $ \nu \in \{ 2,\ldots,d\}$.

The general approach to solution of coupled non-linear 
Eqs.(\ref{vitesse}),(\ref{systemeh1}) and (\ref{systemeh2})
has been discussed in detail in  Ref.\cite{benichouPRB}. Here we merely note that 
despite the fact that using the decoupling scheme in
Eq.(\ref{decouplage}) we effectively close the system of equations on the level of the pair
correlations, solution of Eqs.(\ref{systemeh1}) and (\ref{systemeh2}) (or, equivalently, of  Eqs.(\ref{systemek1}) and
(\ref{systemek2})) still 
poses serious technical 
difficulties: Namely, 
these equations are non-linear with respect to the TP velocity, which enters the gradient term on
the rhs of the evolution equations for the pair correlation, and does depend itself on the values of the
monolayer particles densities in the immediate vicinity of the TP. Solution of this
system of non-linear equation for one- and two-dimensional
substrates is displayed in two next sections.

\section{One-dimensional adsorbed monolayer}\label{1D}

For one-dimensional lattices, which which situation is
 appropriate to adsorbtion on polymer chains \cite{10,11},
general solution of
 Eqs.(\ref{systemeh1}) and (\ref{systemeh2}) has the following form:
\begin{equation}
\label{dprofiles}
k_{n} \equiv k(\lambda) =  \rho_s + K_{\pm} exp\Big(-\sigma |n|/\lambda_{\pm}\Big), 
\;\;\; \lambda = \sigma n, \;\;\; n \in Z,
\end{equation}
where the characteristic 
lengths $\lambda_{\pm}$ obey
\begin{equation}
\label{lambdas}
\lambda_{\pm}= \; \mp \; \sigma \; ln^{-1}\Big[
\frac{A_{1} + A_{-1} + 2 (f + g) \mp \sqrt{\Big(A_{1} + A_{-1} + 2 (f + g)\Big)^2 - 4 A_{1} A_{-1}}}{2 A_{1}}
\Big],
\end{equation} 
while the amplitudes $K_{\pm}$ are given respectively by
\begin{equation}
K_{+} = \rho_s \frac{A_{1} - A_{-1}}{A_{-1} - A_{1} \exp(- \sigma/\lambda_{+})} 
\end{equation}
and
\begin{equation}
K_{-} = \rho_s \frac{A_{1} - A_{-1}}{A_{-1} \exp(- \sigma/\lambda_{-}) - A_{1} } 
\end{equation}
Note that $\lambda_{-} > \lambda_{+}$, and consequently,  the local 
density past the TP approaches its non-perturbed value $\rho_s$
slower than in front of it;  this signifies that
 correlations between the TP position and particle
distribution are stronger past the TP. Next,  $K_{+}$ is always
 positive, while $K_{-} < 0$; this means that 
the density profile is a non-monotoneous function of $\lambda$ 
and is characterized by a jammed region in front
of the TP, in which the local density is higher than $\rho_s$, and a depleted region past the TP in which 
the density is lower than $\rho_s$.

\begin{figure}[h]
\begin{center}
\includegraphics*[scale=0.5]{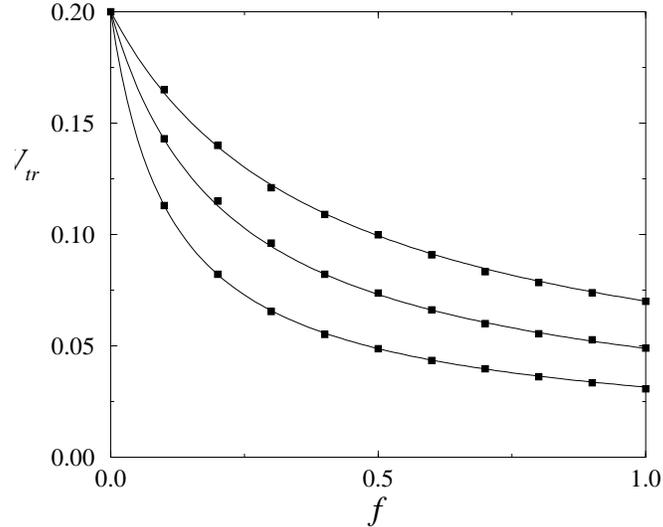}
\end{center}
\caption{\label{vitfig} Terminal velocity of the probe molecule as a function of the adsorption probability $f$
at different values of the parameter $g$. The probe hopping probabilities are $p_1 = 0.6$
and $p_{-1} = 0.4$. The solid lines  give the analytical solution
while the filled squares denote the results of Monte-Carlo simulations. Upper curves correspond to $g = 0.8$, the intermediate -
to $g = 0.5$ and the lower - to $g = 0.3$, respectively.}
\end{figure}  

\begin{figure}[h]
\begin{center}
\includegraphics*[scale=0.5]{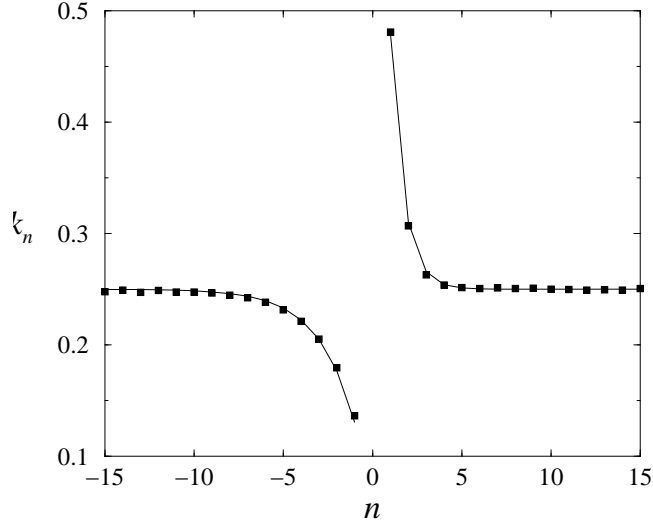}
\end{center}
\caption{\label{proffig} Density profile around stationary moving probe molecule for $f = 0.1$, $g = 0.3$ and $p = 0.98$. 
The solid line is the plot of the analytical solution.
Filled squares are the results of Monte-Carlo simulations.
}
\end{figure}

Now, we are in position to obtain a system of two closed-form non-linear equations determing implicitly the unknown 
parameters $A_{1}$ and $A_{-1}$, which will allow us to compute
 the TP terminal velocity, related to $A_{\pm 1}$ through 
$V_{tr} = \sigma (A_{1} - A_{-1})/2 \tau^*$.
Substituting Eq.(\ref{dprofiles}) into Eq.(\ref{defA}), we find 
\begin{equation}
\label{o}
A_{1} = 1 + \frac{p_1 \tau^*}{\tau} \Big[1 - \rho_s - \rho_s \frac{A_{1} - A_{-1}}{A_{-1} exp(\sigma/\lambda'_{+}) - A_{1}}
\Big]
\end{equation}
and 
\begin{equation}
\label{b}
A_{-1} = 1 + \frac{p_{-1} \tau^*}{\tau} \Big[1 - \rho_s - \rho_s \frac{A_{1} - A_{-1}}{A_{-1}  - A_{1} exp(\sigma/\lambda'_{-})}
\Big]
\end{equation}
Resolution of this system leads then to the stationary
velocity of the TP (see Fig.\ref{vitfig}) as well as the density profiles (see
Fig.\ref{proffig}). 
For arbitrary values of $p$, $f$ and $g$  the parameters $A_{\pm 1}$, defined by
Eqs.(\ref{o}) and (\ref{b}), and consequently, the terminal velocity $V_{tr}$ can be determined
only numerically (see Figs.2 to 4).  However, $V_{tr}$ can be found analytically in the explicit form 
in the limit of
a vanishingly  small force $E$, $E \to 0$. 
Expanding $A_{\pm 1}$ in the Taylor series in powers of $E$ and retaining
only linear with $E$ terms, we find that
the TP velocity follows 
\begin{equation}
\label{st}
V_{tr} \sim \zeta^{-1} E,
\end{equation}
which relation can be thought off as the analog of the Stokes formula for driven motion in a one-dimensional adsorbed monolayer 
undergoing continuous particles exchanges with the vapor phase. Equation (\ref{st}) signifies that the frictional force exerted on the TP by the monolayer
particles is $viscous$. The friction coefficient, i.e. the proportionality factor in Eq.(\ref{st}) is given explicitly by
\begin{equation}
\label{dfriction}
\zeta = \frac{2 \tau}{\beta \sigma^2 (1 - \rho_s)} \Big[1 + \frac{\rho_s \tau^*}{\tau (f + g)}
\frac{2}{1 + \sqrt{1 + 2 (1 + \tau^* (1 - \rho_s)/\tau)/(f + g)}} \Big]
\end{equation}
Note that the friction
coefficient in Eq.(\ref{dfriction}) can be written down as the sum of two contributions $
\zeta=\zeta_{cm}+\zeta_{coop}$. The first one, $\zeta_{cm}=2\tau/ \beta \sigma^2(1-\rho_s)$
is a typicial mean-field result and corresponds to a perfertly
homogeneous monolayer (see discussion following Eq.(\ref{defk})). The second
one,
\begin{equation}
\zeta_{coop}=\frac{8\tau^*\rho_s}{\beta\sigma^2(1-\rho_s)(f+g)}\frac{1}{1+\sqrt{1 + 2 (1 + \tau^* (1 - \rho_s)/\tau)/(f + g)}},
\end{equation} 
has, however, a more complicated
origin. Namely,  it reflects a cooperative behavior
emerging in the monolayer, associated with the formation of
inhomogeneous density profiles (see Fig.\ref{proffig}) - the formation of a ``traffic jam'' in front of
the TP and a ``depleted'' region  past the TP (for more details, see  \cite{benichou}).  The characteristic lengths of these two
regions as well as the amplitudes $K_{\pm}$ depend on the magnitude of the TP velocity; on the other hand, the TP velocity is itself
dependent on the density profiles, in virtue of Eq.(\ref{vitesse}). This results in an intricate interplay between the jamming
effect of the TP and smoothening of the created inhomogeneities by diffusive processes. Note also that cooperative behavior becomes
most prominent in  the conserved particle number limit \cite{bur1,bur2}. Setting $f,g \to 0$, while keeping their ratio fixed (which
insures that $\rho_s$ stays constant), one notices that $\zeta_{coop}$ gets infinitely large. As a matter of fact, as it has been
shown in Refs.\cite{bur1} and \cite{bur2}, in such a situation no stationary density profiles around the TP exist; the size of both
the "traffic jam" and depleted regions grow in proportion to the TP mean displacement $\overline{X_{tr}(t)} \sim \sqrt{t}$
\cite{bur1,bur2}. Consequently, in the conserved particle number limit $\zeta_{coop}$ grows indefinitely in proportion to $\sqrt{t}$.

In order to check
our analytical predictions, we have  performed numerical Monte
Carlo simulations of the exact Master equation of the problem, using the method of
Gillespie \cite{gillespie}. Results of these simulations, performed at different values
of the parameters $f$, $g$, and $p_1$, are also represented in
Figs.\ref{vitfig} and \ref{proffig}.

\begin{figure}[h]
\begin{center}
\includegraphics*[scale=0.5]{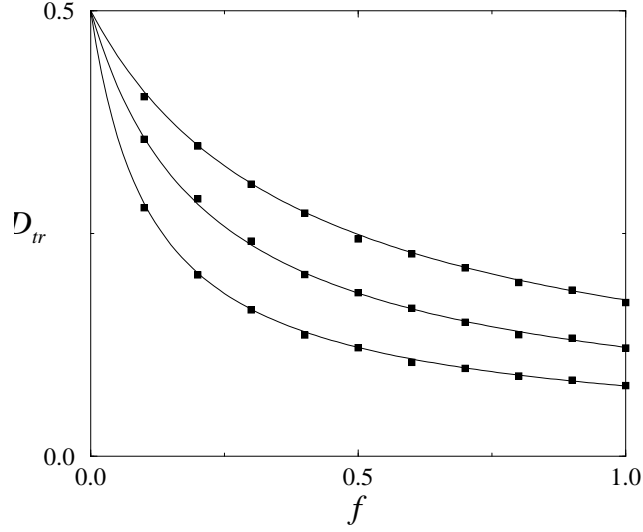}
\end{center}
\caption{\label{self}Self-diffusion coefficient of the probe molecule as a function of the adsorption probability $f$. 
Notations and values of
$g$ are the same as in Figs.\ref{vitfig}.}
\end{figure}

Consider finally the situation with $E = 0$, in which case the terminal velocity vanishes 
and one for which one expects conventional diffusive motion with the mean square displacement of the form
\begin{equation}
\label{diff}
\overline{X_{tr}^2(t)} =  2 D_{tr} t,
\end{equation}
where $D_{tr}$ is some unknown function of the system parameters. Heuristically,  we can compute  $D_{tr}$ 
for the system under study
if we assume the validity 
of the Einstein relation $ D_{tr}=\beta/ \zeta$ between the friction
coefficient 
and the self-diffusion 
coefficient $D_{tr}$ of the TP 
\cite{lebowitz2}, which yields
\begin{equation}
\label{ddiffusion}
D_{tr} = \frac{\sigma^2 (1 - \rho_s)}{2 \tau} \left\{1 +\frac{\rho_s\tau^*}{\tau(f+g)} 
\frac{2}{1 + \sqrt{1 +2 (1 + \tau^* (1 - \rho_s)/\tau) /(f+g)}} \right\}^{-1}.
\end{equation}
 Monte Carlo
simulations (see Fig.\ref{self}) of the system evidently
confirm our prediction for $D_{tr}$ given by Eq.(\ref{ddiffusion}), and hence, confirm the 
validity of the Einstein relation for the system under study. This is, of course, not an unexpected, but still 
a non-trivial result \cite{lebowitz2}.

\section{Two-dimensional adsorbed monolayer}\label{2D}

We turn now to the case of a two-dimensional substrate - adsorbtion onto
the surface of a solid exposed to a vapor phase. Here, the situation gets somewhat more difficult from the computational
 point of view;  we have now to
solve the partial difference equations problem (\ref{systemeh1}),
(\ref{systemeh2}) rather than the mere difference
equations arising in the one-dimensional case.  

Solution in two-dimensions can be found in a most convenient fashion if we introduce 
the
generating function for the particle density profiles, defined as
\begin{equation}
H(w_1,w_2)\equiv\sum_{n_1=-\infty}^{+\infty}\sum_{n_2=-\infty}^{+\infty}h_{n_1,n_2}w_1^{n_1}w_2^{n_2},
\label{defH}
\end{equation}
where $h_{n_1,n_2}\equiv h(n_1{\bf e_1}+n_2{\bf e_2})$. 
Multiplying both sides of Eqs. (\ref{systemeh1}) and (\ref{systemeh2}) by $w_1^{n_1}w_2^{n_2}$,
 and performing summations over $n_1$ and $n_2$,
we find that $H(w_1,w_2)$ is  given explicitly by
\begin{eqnarray}
H(w_1,w_2)&=& - \; K(w_1,w_2)\;\Big\{A_1w_{1}^{-1}+A_{-1}w_1+A_2(w_2+w_2^{-1})-\alpha\Big\}^{-1}, 
\label{adevelopper}
\end{eqnarray}
where
$
\alpha\equiv\sum_\nu A_\nu+4(f+g)$
and 
\begin{equation}
K(w_1,w_2)\equiv \sum_\nu
A_\nu(w_{|\nu|}^{\nu/|\nu|}-1)h({\bf e}_{\boldsymbol
\nu})+\rho_s(A_1-A_{-1})(w_{1}-w_1^{-1})
\label{defK}. 
\end{equation}
Equations (\ref{adevelopper}) and (\ref{defK}) determine the
generation function for the density profiles exactly.

Before we proceed to the inversion of $H(w_1,w_2)$ with respect to the variables $w_1$ and $w_2$, we note that
we can access 
interesting integral characteristic of the density profiles directly using the result in  Eqs.(\ref{adevelopper}) 
and (\ref{defK}). 
Namely, as we have already remarked,  the presence of the driven TP
 induces
 an inhomogeneous density distribution
in the monolayer. One can thus pose a natural question whether 
equilibrium between adsorption and desorption processes gets shifted due to such a perturbancy, 
i.e. whether 
the equilibrium density in the monolayer is different from that given by Eq.(\ref{rhostat}). The answer is trivially "no"
 in the case when the particles number is
explicitly conserved, but in the general case 
with arbitrary $f$ and $g$ this is not at all evident: similarly to the behavior in one-dimensional system 
one expects that also in two-dimensions the density  profiles are asymmetric as seen from the stationary moving 
TP and are characterized by a
condensed, "traffic-jam"-like region in front of and a depleted region past the TP. One anticipates then that the
desorption events are 
favored in front of the TP, while  the adsorption events are evidently suppressed by the excess density.
On the other hand, past the TP desorption is diminished due to 
the particles depletion while adsorption
may proceed more readily due to the same reason.
 It is thus not at all clear ${\it a \; priori}$ whether these two
effects can compensate each other exactly, in view of a possible
asymmetry of the density profiles, as it happens in the
one-dimensional model (see Fig.\ref{proffig}).

For this purpose, we study the behavior of  the integral
deviation $\Omega$ of the density from the equilibrium value $\rho_{s}$, i.e.
$\Omega\equiv\sum_{n_1=-\infty}^{+\infty}\sum_{n_2=-\infty}^{+\infty}h_{n_1,n_2}$,
which can be computed straightforwardly from
 Eqs.(\ref{adevelopper}) and  (\ref{defK}) by setting both $w_1$ and $w_2$ equal to unity.  Noticing that
$K(w_1=1,w_2=1)=0$, and that
$A_1+A_{-1}+2A_2-\alpha=-4(f+g)$, i.e. is strictly negative as soon as 
adsorption/desorption processes are present, we obtain then
that $\Omega$ is stricly equal to 0.  This implies, in turn,  that the 
perturbancy of the density distribution in the monolayer created by the driven TP
does not shift the global balance between the adsorption and desorption events. 

Inversion of the generating function with respect to $w_1$ and $w_2$ 
requires quite an involved mathematical analysis, which has been presented in detail in
 Ref.\cite{benichouPRB}. General solution
for the density profiles reads:
\begin{eqnarray}
h_{n_1,n_2}=
\alpha^{-1}\Big\{\sum_\nu A_\nu h({\bf e}_{\boldsymbol \nu})\nabla_{-\nu}F_{n_1,n_2}-
\rho_s(A_1-A_{-1})(\nabla_1-\nabla_{-1})F_{n_1,n_2}\Big\}.
\label{2dsolh}
\end{eqnarray}
with
\begin{eqnarray}
F_{n_1,n_2}=
\left(\frac{A_{-1}}{A_1}\right)^{n_1/2}\int_0^{\infty}e^{-t}{\rm I}_{n_1}\left(2\alpha^{-1}\sqrt{A_1A_{-1}}t\right){\rm I}_{n_2}\left(2\alpha^{-1}A_2t\right){\rm d}t,
\label{repint}
\end{eqnarray}
where ${\rm I}_n(z)$ stands for the  modified Bessel function. We mention that
$F_{n_1,n_2}$  has   an interesting physical interpretation in terms of
the generating function of a random walk of a single particle (that
is, in absence of the particles environment) \cite{hughes}.

Now, the Eqs.(\ref{2dsolh}) and (\ref{repint}) display $h_{n_1,n_2}$
as a function of the coefficients $A_\nu$ that remain to be determined.
As a matter of fact, these coefficients depend  themselves
on the local densities in the immediate vicinity of the tracer, i.e. on $h({\bf e_\nu})$. 
This implies that we have to determine them  from Eqs.(\ref{2dsolh}) 
and (\ref{repint}) in a self-consistent way \cite{benichouPRB}. Some analysis (see \cite{benichouPRB} shows that $A_\nu$ are
determined implicitly as the solution of the following system of three non-linear matrix equations
\begin{equation}
\forall\nu=\{\pm1,2\},\;\;\;\;A_\nu=1+\frac{4\tau^*}{\tau}p_\nu\left\{1-\rho_s-\rho_s(A_1-A_{-1})\frac{\det\tilde{C}_\nu}{\det\tilde{C}}\right\},
\label{2dimplicite}
\end{equation}  
where
\begin{equation}
\tilde{C}=
\begin{pmatrix}
A_1\nabla_{-1}F_{\bf e_1}-\alpha & A_{-1}\nabla_1F_{\bf e_1} & A_{2}\nabla_{-2}F_{\bf e_1}\\
A_1\nabla_{-1}F_{\bf e_{-1}} & A_{-1}\nabla_1F_{\bf e_{-1}}-\alpha & A_{2}\nabla_{-2}F_{\bf e_{-1}}\\
A_1\nabla_{-1}F_{\bf e_2} & A_{-1}\nabla_1F_{\bf e_2} &  A_{2}\nabla_{-2}F_{\bf e_2}-\alpha
\end{pmatrix}.
\end{equation}
the matrix  $\tilde{C_\nu}$ stands for the matrix obtained from  $\tilde{C}$
by replacing the $\nu$-th column  by the column-vector
$\tilde{F}$,
\begin{equation}
\tilde{F}=
\begin{pmatrix}
(\nabla_1-\nabla_{-1})F_{\bf e_1}\\
(\nabla_1-\nabla_{-1})F_{\bf e_{-1}}\\
(\nabla_1-\nabla_{-1})F_{\bf e_2}
\end{pmatrix},
\end{equation}
while the local deviations $h({\bf e}_{\boldsymbol \nu})$ are expressed in terms of $A_\nu$ as
\begin{equation}
h({\bf e}_{\boldsymbol \nu})=(1-\rho_s)+\frac{\tau}{4\tau^*p_\nu}(1-A_\nu)
\label{ppasnul}
\end{equation}  
Lastly, the TP terminal velocity obeys
\begin{equation}
V_{tr}=\frac{\sigma}{\tau}(p_1-p_{-1})(1-
\rho_s)\Big\{1+\rho_s\frac{4\tau^*}{\tau}\frac{p_1\det\tilde{C}_1-p_{-1}\det\tilde{C}_{-1}}{\det\tilde{C}}\Big\}^{-1},
\label{forcevitesse}
\end{equation}
which represents the 
desired general force-velocity relation 
for the system under study, which is valid for arbitrary
magnitude of the external bias and arbitrary values of 
other system's parameters. 

\subsection{Asymptotical behavior of the density profiles at large separations from the TP.}

Asymptotical behavior of the density profiles at large distances from
the TP follows from the analysis of the analyticity properties of the complex function
$N(z)\equiv\sum_{n=-\infty}^{+\infty}h_{n,0}z^n$. It has been  shown in  Ref.\cite{benichouPRB}
that in front of the TP,
the deviation $h_{n,0}$ always decays exponentially with the distance:
\begin{equation}
h_{n,0}\sim K_+\frac{\exp\Big(-n/ \lambda_+\Big)}{n^{1/2}},
\end{equation}
where the characteristic length $\lambda_+$ obeys:
\begin{equation}
\lambda_+ =
\ln^{-1}\Big(\frac{1}{A_{-1}}\left\{\frac{\alpha}{2}-A_2+\sqrt{\left(\frac{\alpha}{2}-A_2\right)^2-A_1A_{-1}}\right\}\Big)
\end{equation}
Note that $\lambda_+$ stays finite for any values of the system parameters.

On contrary, the behavior of the density profiles at large
distances past the tracer qualitatively  depends on the physical
situation studied. In the general case when  exchanges with the
particles reservoir are taken into allowed, the decay of the density
profiles is still exponential with the distance:
\begin{equation}
h_{-n,0}\sim K_-\frac{\exp\Big(-n/ \lambda_-\Big)}{n^{1/2}},
\end{equation}    
where
\begin{equation}
\lambda_- = -
\ln^{-1}\Big(\frac{1}{A_{-1}}\left\{\frac{\alpha}{2}-A_2-\sqrt{\left(\frac{\alpha}{2}-A_2\right)^2-A_1A_{-1}}\right\}\Big)
\end{equation}
Note that in the general case the characteristic lenghts again, similarly to the one-dimensional case,
satisfy the inequality $\lambda_->\lambda_+$, which means that the
correlations between the TP  and the particles of the monolayer are
always stronger past than in front of the TP. 

Such correlations can
even become extremely strong in the special case when the particles
exhanges with the vapor phase are forbidden, i.e. in the conserved particles number limit, which can be realized for the monolayers
sandwiched in a narrow gap between two solid surfaces. In this case, we have that $\lambda_-$ becomes infinitely large 
and, in the limit $n\to+\infty$, 
the deviation of the particle density from the equilibrium value $\rho_s$ follows
\begin{equation}
h_{-n,0}=-\frac{K_-'}{n^{3/2}}\left(1+\frac{3}{8n}+{\mathcal O}\Big(\frac{1}{n^2}\Big)\right),
\label{algebrique}
\end{equation} 
Remarkably enough, in this case the correlations between the TP position
and the particles distribution
vanish  {\it algebraically} slow with the distance! This implies, in turn,
 that in the conserved particles
number case, the mixing of the monolayer is not efficient enough to prevent the appearence of the quasi-long-range order
and the medium "remembers"
 the passage of the TP
on a long time and space scale, which signifies very 
strong memory effects.

\subsection{Limit of small applied force}

We turn now to the limit $\beta E \ll 1$, in which case the problem simplifies considerably and allows to obtain explicit results
for the local densities in the immediate vicinity of the TP and consequently, for the TP terminal velocity and
its diffusivity.

In this limit, we arrive again at a Stokes-type formula of the form $V_{tr}\sim E/ \zeta $, where now
\begin{equation}
\label{rrr}
\zeta=\frac{4\tau}{\beta\sigma^2(1-\rho_s)}\left\{1
+\frac{\tau^*}{\tau}\frac{\rho_s}{\Big(f + g + 1 + \tau^* (1 - \rho_s)/\tau\Big)\Big({\cal L}(x)-x\Big)}\right\},
\end{equation} 
with
\begin{equation}
x = \frac{1}{2} \frac{1 + \tau^* (1 - \rho_s)/\tau}{f + g + 1 + \tau^* (1 - \rho_s)/\tau},
\end{equation}
and
\begin{equation}
{\cal L}(x)\equiv\left\{\int_0^\infty e^{-t}\Big(({\rm I}_0(xt)-{\rm I}_2(xt)){\rm I}_0(xt){\rm d}t\right\}^{-1}.
\end{equation}
Note  that we again are able to sigle out two physical meaningful contributions to 
the friction coefficient $\zeta$.  Namely, the first term on the rhs of Eq.(\ref{rrr}) is just the mean-field-type result
corresponding to a perfectly stirred monolayer, in which correlations between the TP and the monolayer particles are discarded.
The second term, similarly to the one-dimensional case, mirrors the cooperative behavior emerging in the monolayer
and is associated with the backflow effects.  In contrast to the one-dimensional case, however, the contribution to the overall
friction coefficient stemming out of the cooperative effects remains finite in the conserved particles limit.   

We also wish to remark,
that a qualitatively similar physical effect has been predicted recently
for a  different model system involving a charged particle moving 
at a constant speed a small distance above the surface of an incompressible,
 infinitely deep liquid. It has been shown in Refs.\cite{elie1,elie2}, 
that the interactions between the moving particle and the fluid molecules
induce an effective frictional force exerted on the particle, producing
a local distortion of the liquid interface, - a bump, which travels 
together with the particle and increases effectively its mass.  The mass of the bump, 
which is
analogous to the jammed region appearing in our model, depends itself
on the particle's velocity resulting
in a non-linear coupling  between the medium-induced
 frictional force exerted on the particle and its
velocity \cite{elie1,elie2}.

Lastly, assuming {\it a priori} that the Einstein relation holds for the system
under study, we estimate the TP diffusion coefficient $D_{tr} = \beta^{-1}\zeta^{-1}$ as
\begin{equation}
D_{tr}=\frac{\sigma^2}{4\tau}(1-\rho_s)\left\{1-\frac{2\rho_s\tau^*}{\tau}\frac{1}{ 4 
(f + g + 1 + \tau^* (1 - \rho_s)/\tau) {\cal L}(x)-1+(3\rho_s-1)\tau^*/\tau}\right\}.
\label{2dautodiffgen}
\end{equation}
It seems now interesting to compare our general result in Eq.(\ref{2dautodiffgen}) against the classical result of Nakazato and
Kitahara \cite{nakazato}, which describes TP
 diffusion coefficient in a two-dimensional lattice gas with conserved
particles number. 
Setting $f$ and $g$ equal to zero, while assuming that their ratio 
has a  fixed value, $f/g=\rho_s/(1-\rho_s)$, we have then 
that 
\begin{equation}
\label{d}
\hat{D}_{tr}=\frac{\sigma^2}{4\tau}(1-\rho_s)\left\{1-\frac{2\rho_s\tau^*}{\tau}\frac{1}{4 (1 + \tau^* (1 - \rho_s)/\tau) {\cal L}(1/2)-1+(3\rho_s-1)\tau^*/\tau}\right\}.
\end{equation}
Using next the equality \cite{mccrea}:
\begin{equation}
{\cal L}(1/2)=\frac{\pi}{4 (\pi - 2)},
\end{equation}
we find that the right-hand-side of Eq.(\ref{d}) attains the form
\begin{equation}
\label{comparNaka}
\hat{D}_{tr}=\frac{\sigma^2}{4\tau}(1-\rho_s)\left\{1-
\frac{2\rho_s\tau^*}{\tau}\frac{1-2/\pi}{1+(1-\rho_s)\tau^*/\tau-(1-2/\pi)(1+(1-3\rho_s)\tau^*/\tau)}\right\},
\end{equation}
which expression
 coincides exactly with the earlier result obtained  in 
Refs.\cite{nakazato} and \cite{tahir} within the framework of a different, compared to ours,
 analytical techniques. The result in Eq.(\ref{comparNaka}) is
known to be exact in the limits $\rho_s \ll 1$ and $\rho_s \sim 1$, and serves as a very good approximation for the
self-diffusion coefficient in hard-core lattice gases of arbitrary density \cite{kehr}, which 
supports in a way the
validity of the
approximation invoked in Eq.(\ref{decouplage}).

\section{Conclusion}\label{Conclusion}

To conclude,  we have studied analytically the
 intrinsic frictional properties
 of adsorbed monolayers,
composed of mobile hard-core particles  
undergoing continuous exchanges with the vapor. 
Our analytical approach has been based on
the master equation, describing the time
 evolution of the system, 
which has allowed us to evaluate a system of coupled
dynamical equations 
for the TP
velocity and a 
hierarchy of correlation functions. 
To solve these coupled equations, we have invoked an approximate closure scheme
based on the decomposition of the
third-order correlation functions into a product of pairwise correlations, which has
been  
first 
introduced in Ref.\cite{bur1} for  a related
 model of a driven tracer particle dynamics in a one-dimensional lattice gas 
with conserved particles number. 
Within the framework of this approximation,
we have derived a system of coupled, discrete-space equations describing evolution 
of the density profiles in the adsorbed monolayer, as seen from the  moving
tracer, and its velocity $V_{tr}$.  We have shown that  the density profile  around the tracer is strongly
inhomogeneous: the local density of the adsorbed
 particles in front of the tracer is higher than the 
average and approaches the average value as an exponential
 function of the distance from the tracer. 
On the other hand, past the tracer 
the local density is always lower than the average, and depending on
whether the number of particles 
is explicitly conserved or not, the local density past the tracer
 may tend to the average value either as an exponential or even as an
 $\it algebraic$ function of the distance. The latter reveals 
especially strong memory effects and strong 
correlations between the particle distribution in the
environment and the carrier position. 
Next, we have derived a general force-velocity relation, which  defines the terminal velocity of the tracer particle
for arbitrary applied fields and arbitrary values of other system parameters. 
We have demonstrated next that in the limit of a vanishingly small external bias this relation attains a simple, but physically
meaningful form of the Stokes formula, which signifies that in this limit the frictional force exerted on the tracer  by
the adsorbed monolayer particles is viscous. Corresponding friction coefficient has been also explicitly determined. 
In addition, we estimated the self-diffusion
coefficient of the tracer in the absence of the field and showed that it reduces to the well-know result of
Refs.\cite{nakazato} and \cite{tahir} in the limit when the particles number is conserved.

\end{document}